\documentclass[twoside,leqno,twocolumn]{article}

\usepackage[letterpaper]{geometry}
\usepackage{enumitem}
\usepackage{booktabs}
\usepackage{makecell}
\usepackage{tabulary}
\usepackage{amsmath,amsfonts,amssymb}
\usepackage{subcaption}
\usepackage{multirow}
\usepackage{multicol}
\usepackage{url}
\usepackage{ltexpprt}
\usepackage{graphicx}
\usepackage[export]{adjustbox}
\usepackage{xcolor}
\usepackage{bm}

\newtheorem{mydef}{Definition}
\usepackage{cleveref}
\crefname{equation}{Eq.}{Eq.}
\crefname{section}{Sec.}{Sections}
\crefname{subsection}{Sec.}{Sections}
\crefname{subsubsection}{Sec.}{Sections}
\crefname{figure}{Fig.}{Figures}
\crefname{table}{Table}{Tables}
\crefname{subfigure}{Fig.}{Figures}
\crefname{algocf}{Algorithm}{Algorithms}

\begin{document}
\title{BasConv: Aggregating Heterogeneous Interactions for Basket Recommendation with Graph Convolutional Neural Network}
\author{Zhiwei~Liu\thanks{work is done during internship at Walmart Labs}~\thanks{University of Illinois at Chicago;\{zliu213, psyu\}@uic.edu}
\and Mengting Wan\thanks{Airbnb, Inc; mengting.wan@airbnb.com; Work was done while the author was at UC San Diego.}
\and Stephen Guo\thanks{Walmart Labs; \{sguo,kachan\}@walmartlabs.com}
\and Kannan Achan\footnotemark[4]
\and Philip S. Yu\footnotemark[1]
}

\date{}

\maketitle
\fancyfoot[R]{\scriptsize{Copyright \textcopyright\ 2020 by SIAM\\
Unauthorized reproduction of this article is prohibited.}}

\begin{abstract}
    Within-basket recommendation reduces the exploration time of users, where the user's intention of the basket matters. The intent of a shopping basket can be retrieved from both user-item collaborative filtering signals and multi-item correlations. By defining a basket entity to represent the basket intent, we can model this problem as a basket-item link prediction task in the User-Basket-Item~(UBI) graph. Previous work solves the problem by leveraging user-item interactions and item-item interactions simultaneously. However, collectivity and heterogeneity characteristics are hardly investigated before. Collectivity defines the semantics of each node which should be aggregated from both directly and indirectly connected neighbors. Heterogeneity comes from multi-type interactions as well as multi-type nodes in the UBI graph. To this end, we propose a new framework named \textbf{BasConv}, which is based on the graph convolutional neural network. Our BasConv model has three types of aggregators specifically designed for three types of nodes. They collectively learn node embeddings from both neighborhood and high-order context. Additionally, the interactive layers in the aggregators can distinguish different types of interactions. Extensive experiments on two real-world datasets prove the effectiveness of BasConv. Our code is available online\footnote{\url{https://github.com/JimLiu96/basConv}}.

\end{abstract}
\section{Introduction}
Shopping for a group of items during a session is a common behavior of users when shopping online~\cite{le2017basket,wan2018representing,kang2019complete}. We define a \textit{basket} to contain a set of items which are bought at the same time by the same user~\cite{le2017basket,bai2019personalized,xu2019modeling}. \textit{Within-basket recommendation}~\cite{le2017basket,wan2018representing} is to recommend items for a shopping basket, which can reduce the exploration time of users. In order to make a recommendation, we need to understand the \textit{intent} of the basket. For example, \textit{milk} can be either more related to \textit{bread} or to \textit{flour} when the purpose of the corresponding basket is for \textit{breakfast} or for \textit{making a cake}, respectively. Ignoring the intent of a basket compromises the ability of a model to distinguish the different item relationships within the same basket. The intent of the basket should be retrieved from two perspectives: 
(1) \textit{User-item collaborative filtering} (CF) signals, which model item semantic information as well as the user's personal tastes~\cite{rendle2009bpr,he2017neural,rendle2010factorization} and (2) \textit{multi-item correlations}, which reveal users' intents for a shopping basket~\cite{xu2019modeling,bai2019personalized,wan2018representing,mcauley2015inferring}, e.g. \{\textit{milk}, \textit{flour}, \textit{sugar}, \textit{egg}\} contributes to the purpose of making a cake.

\begin{figure}
    \centering
    \includegraphics[width=0.95\linewidth]{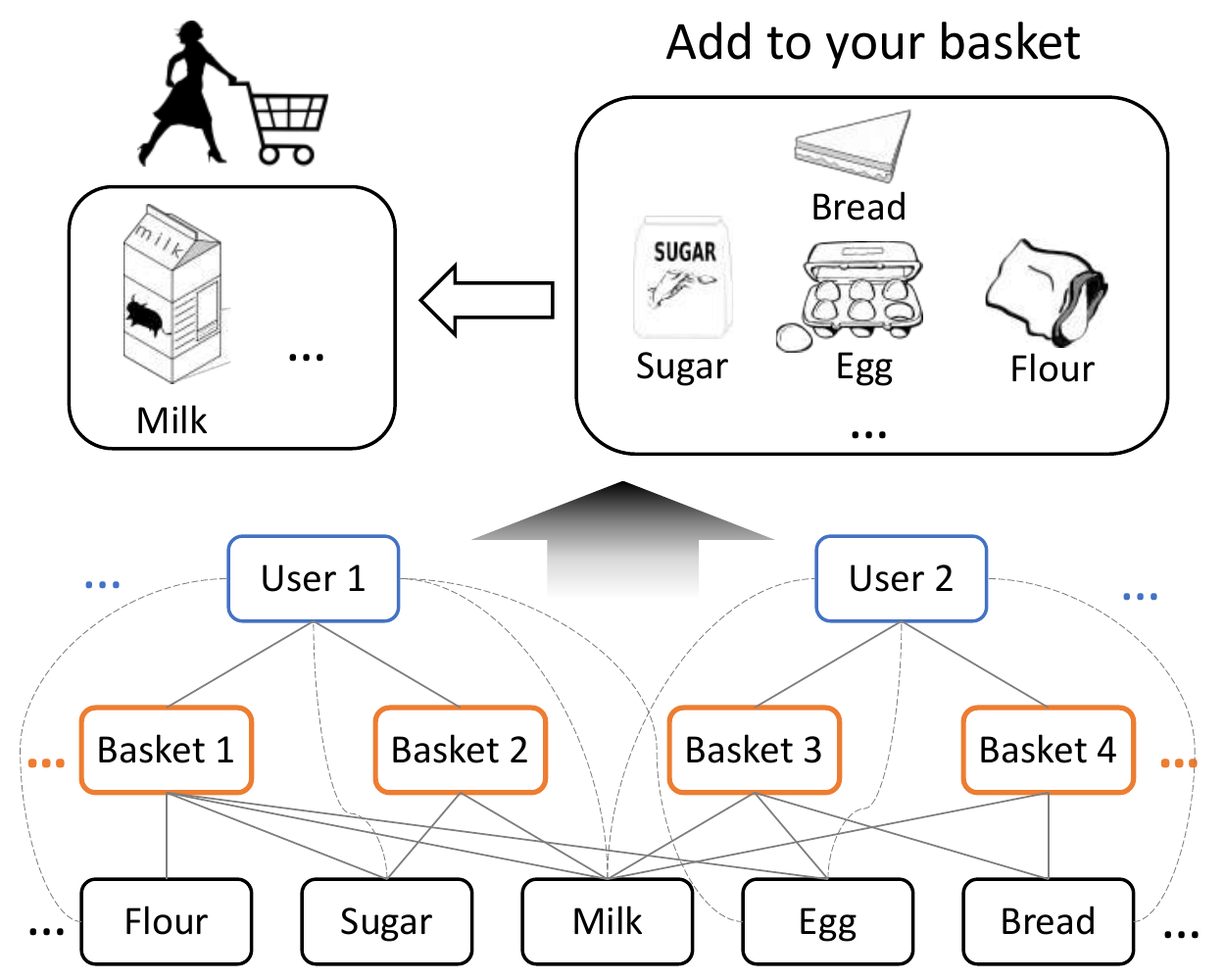}
    \caption{The top figure shows an example basket recommendation for when milk is in the basket, we recommend items to the current basket. The bottom figure is an example of a UBI graph where user-basket, basket-item, and user-item interactions exist.}
    \label{fig:illustration}
\end{figure}

As illustrated in \cref{fig:simple_ubi}, we introduce the \textbf{user-basket-item (UBI) graph} to characterize these two types of structures simultaneously. Unlike the traditional user-item bipartite graph \cite{zheng18spectral,wang19neural,hamilton2017inductive}, \textit{basket} nodes are incorporated to represent the semantics of users' shopping baskets. The basket recommendation problem can thus be defined as predicting the links between basket nodes and item nodes.  
On top of this graph, we propose a new framework named \textbf{BasConv} to tackle this problem with a graph convolutional neural network~(GCNN)~\cite{kipf17semi,wang19neural,liu2019jscn}. Different from prior work, we are able to address the following two aspects:

\begin{itemize}
    \item \textbf{High-Order Collectivity.} The semantic information of each node can be collected from its neighbors and other relevant nodes through high-order paths. For example, to represent a basket node, information can be aggregated from not only the associated user and the items included in this basket, but also other shopping baskets owned by the same user. In this way, interactions among users, baskets, and items can be holistically modeled.
    \item \textbf{Heterogeneity.} As shown in \cref{fig:ubi-aggregator}, by differentiating the types of nodes (i.e.,~user, basket or item), we build different aggregators to propagate information on the UBI graph. By doing so, heterogeneous relationships (e.g. user-basket and item-basket interactions) can be distinguished. 
\end{itemize}
Specifically, in our proposed framework BasConv, the user aggregator (\cref{fig:user-agg}) generates the personalized user embeddings by aggregating the information of all connected baskets and items. The basket aggregator (\cref{fig:basket-agg}) summarizes the multi-item correlations inside the corresponding baskets and combines it with the associated user. The item aggregator (\cref{fig:item-agg}) yields item embeddings by collecting the intents of their corresponding baskets.
Moreover, the recursive learning procedure and the multi-layer structure of BasConv capture the high-order information over the UBI graph. 
The contributions of this paper are as follows:

\begin{itemize}
    \item \textbf{New Framework:} We propose a new framework, BasConv, to tackle the basket recommendation problem. We model the recommendation task as predicting the interactions over the UBI graph. Then, we design heterogeneous aggregators to learn the embedding of each node. Finally, the predictive layer outputs the ranking results from the learned embeddings. 
    
    \item \textbf{Heterogeneous Interactions:} Various interactive layers in BasConv retrieve the heterogeneous interactions on the UBI graph. The interactive layers explicitly model the user-item, basket-item, and user-basket interactive signals. 
    
    \item \textbf{Heterogeneous Aggregators:} We design three different aggregators for user, basket and item entities, which are built upon heterogeneous interactive layers to retrieve both heterogeneous nodes and heterogeneous linkage signals in the UBI graph. The same type of aggregators in the same layer share common training parameters.
    
\end{itemize}{}

\section{Related Work}
In this section, we review two research areas: basket recommendation and graph convolutional neural network (GCNN)-based recommender systems. 
\subsection{Basket Recommendation.}
Basket recommendation requires not only the user-item CF signals~\cite{he2017neural,zheng18spectral,wang19neural}, but also the item-item relationships~\cite{mcauley2015inferring,wan2018representing}, e.g., the complementary and substitution relationships. Item A is a substitute for item B if A can be purchased instead of B,
while item A is complementary to item B if it can be purchased in addition
to B~\cite{mcauley2015inferring}. Both of these concepts are extensively investigated in the previous work~\cite{mcauley2015inferring,wan2018representing,xu2019modeling}. Sceptre~\cite{mcauley2015inferring} is proposed to model and predict relationships between items
from the text of their reviews and the corresponding descriptions. Item2vec~\cite{barkan2016item2vec} learns the item embedding from the user-generated item sets, i.e. the baskets, based on the word2vec model. BFM~\cite{le2017basket} learns one more basket-sensitive embedding for each item rather than only one embedding, which can help to find item relation w.r.t. the current shopping basket. Prod2vec~\cite{grbovic2015prod2vec} applies the same idea to learn the distributed representation of items and support the recommendation of the ad in \textit{Yahoo! Mail}. Triple2vec~\cite{wan2018representing} improves the within-basket recommendation via (user, item A, item B) triplets sampled from baskets, where item A and item B have a complementary relationship. Later, ~\cite{xu2019modeling} proposed a Bayesian network to unify the context information and high-order relationships of items, learning context-aware dual embeddings of items. We argue that these works have no discrimination towards heterogeneous types of interactions and explore little on the collectivity pattern in the basket recommendation problem.
\begin{figure*}[!hbt]
  \centering
  \begin{subfigure}[b]{0.28\linewidth}
    \centering
    \includegraphics[width=\linewidth]{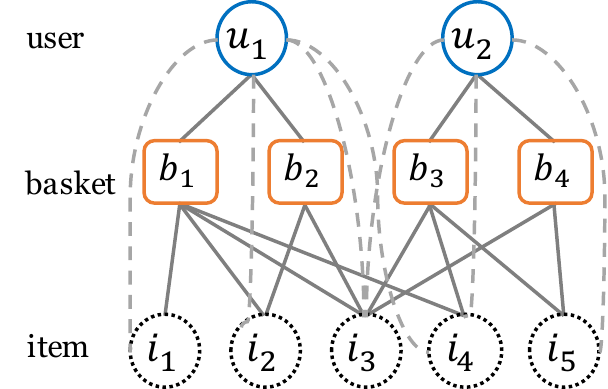}
    \caption{UBI Example}\label{fig:simple_ubi}
  \end{subfigure}  \hfill
  \begin{subfigure}[b]{0.23\linewidth}
    \centering
    \includegraphics[width=\linewidth]{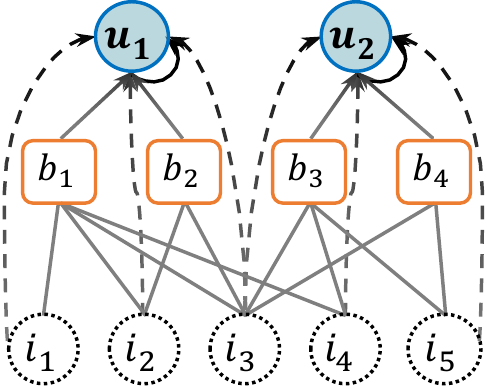}
    \caption{User Aggregator}\label{fig:user-agg}
  \end{subfigure} \hfill
  \begin{subfigure}[b]{0.23\linewidth}
    \centering
    \includegraphics[width=\linewidth]{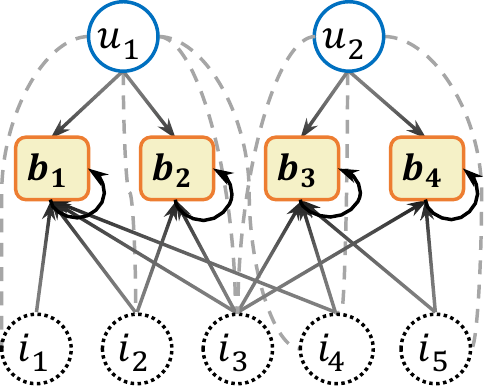}
    \caption{Basket Aggregator}\label{fig:basket-agg}
  \end{subfigure} \hfill
  \begin{subfigure}[b]{0.23\linewidth}
    \centering
    \includegraphics[width=\linewidth]{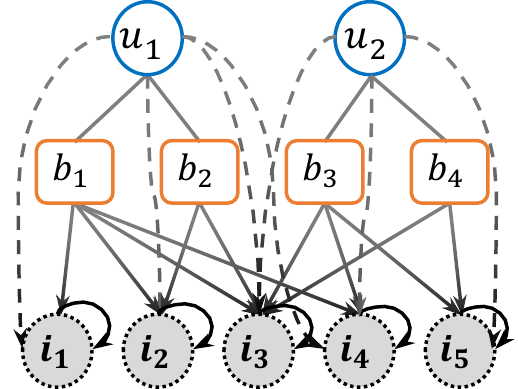}
    \caption{Item Aggregator}\label{fig:item-agg}
  \end{subfigure}
  \hfill
  \vspace{-3mm}
  \caption{We present a UBI graph example in (a). The user, basket, and item aggregator examples are illustrated in (b), (c) and (d), respectively. The aggregated nodes are shadowed with associated colors. 
  }\label{fig:ubi-aggregator}
\end{figure*}
\subsection{GCNN-based Recommender System.}
By defining the user-item interaction as a bipartite-graph~\cite{zheng18spectral,wang19neural}, we can apply recently developed graph convolutional~\cite{kipf17semi,hamilton2017inductive} models to design recommender systems~\cite{berg2017graph,hamilton2017inductive,zheng18spectral}. GCN~\cite{kipf17semi} is proposed to learn the graph embeddings from spectral graph convolutions. Based on this idea, GC-MC~\cite{berg2017graph} predicts the links between users and items by applying the graph convolutional network as the graph encoder.  Graphsage~\cite{hamilton2017inductive} learns the graph embedding by aggregating the information of neighbors, which is extended as a large-scale recommender system, namely PinSage~\cite{pinsage2018ying}. SpectralCF~\cite{zheng18spectral} designs a spectral convolutional filter to model the CF signals in user-item bipartite graphs. NGCF~\cite{wang19neural} explicitly models the high-order CF signal in the user-item bipartite graph. It designs a multi-layer graph convolutional network by constructing, then aggregating the messages over the graph. However, none of the previous methods study the heterogeneity in the graph. Our proposed BasConv model can capture heterogeneous interaction signals. To our best knowledge, BasConv is the first GCNN model to solve the basket recommendation task.

\section{Preliminary and Definition}\label{sec:preliminary}
We have a set of items $I=\{i_{1},i_{2},\dots, i_{|I|}\}$ and a set of users $U=\{u_{1}, u_{2}, \dots, u_{|U|}\}$. Given a partial
shopping basket $b$ which contains a set of items $I_{b} \subset I$ and is associated with a user $u \in U$, we recommend items ($i^{*} \in I\setminus I_{b}$) to complete the current basket $b$. The recommendations are 
based on the intent of the basket, which can be inferred from the semantics of items within $b$ as well as user $u$'s preferences.
In particular, we define a user-basket-item (UBI) graph to represent
interactions among users, baskets and items. 
\begin{mydef}
\textbf{(UBI Graph).}
A UBI graph is defined as $\mathcal{G}=$ $ (\mathcal{V}_u, \mathcal{V}_b, \mathcal{V}_i, \mathcal{E}_{ub}, \mathcal{E}_{bi}, \mathcal{E}_{ui} )$.  $\mathcal{V}_u$, $\mathcal{V}_b$ and $\mathcal{V}_i$ represent the vertices of user, basket and item, respectively. $\mathcal{E}_{ub}$, $\mathcal{E}_{bi}$ and $\mathcal{E}_{ui}$ denote the edges between users and baskets, between baskets and items and between users and items, respectively. Each basket is 
connected to one user exclusively. 
\end{mydef}
The UBI graph is an extension of the user-item bipartite graph~\cite{zheng18spectral} with one more basket entity. The user-basket part, as illustrated in \cref{fig:simple_ubi}, is of a tree structure, while the user-item and basket-item part are both of bipartite graph structure. We define three different types of interaction matrices from the UBI graph, i.e., $\mathbf{R}_{ub}$, $\mathbf{R}_{bi}$ and $\mathbf{R}_{ui}$ for the user-basket interaction matrix, basket-item interaction matrix, and user-item interaction matrix, respectively. We show the user-basket interaction matrix as

\begin{equation}\label{eq:feedback matrix}
\mathbf{R}_{ub}(r,j) = \left\{\begin{matrix}
1 & \text{if $(u_r, b_j) \in \mathcal{E}_{ub}$} \\
0 & \text{otherwise}.
\end{matrix}\right.
\end{equation}

We define the other two interaction matrices $\mathbf{R}_{bi}$ and $\mathbf{R}_{ui}$ in the same way, as in Eq.~(\ref{eq:feedback matrix}) from the UBI graph with entity substitution, i.e. $\mathbf{R}_{bi}(j,k)=1$ when $(b_j, i_k) \in \mathcal{E}_{bi}$ and $\mathbf{R}_{ui}(r,k)=1$ when $(u_r, i_k) \in \mathcal{E}_{ui}$.  

\section{BasConv Model}\label{sec:basconv}
In this section, we present the structure of our proposed BasConv model. BasConv has two major parts, the embedding layer and the heterogeneous aggregators. BasConv has three different types of aggregators, i.e., the basket, user, and item aggregator. The stucture of BasConv is presented in Fig.~\ref{fig:basconv_structure}.
\subsection{Embedding.}
We use $d$ dimensional embedding vectors $e_{u}$, $e_{i}$, $e_{b} \in \mathbb{R}^{d}$ to describe the user, item, and basket entities in the UBI graph, respectively. We can define three embedding matrices to form a look-up table for the embeddings of each type of entity, e.g., the user embedding matrix, $\mathbf{E}_{u} = \left[\mathbf{e}_{u_{1}},\mathbf{e}_{u_{2}},\dots,\mathbf{e}_{u_{|U|}}\right]$
where $|U|$ represents the total number of users in the graph.
The embedding matrices represent the information of the entities in the UBI graph. They are propagated along with the structural information of the UBI graph into the next layer of GCNN. At each layer, we refine the embeddings of all nodes by leveraging both heterogeneous and high-order interactions of the UBI graph. Hence, we use superscript to denote the layer number, e.g., $\mathbf{E}_{u}^{(0)}$ for the initial embedding of users and $\mathbf{E}_{u}^{(l)}$ for the embedding of users at the $l$-th layer. The number of parameters in BasConv at each layer is independent of the number of nodes in the graph and linear to the dimension of embeddings, which will be analyzed in detail later.

\subsection{Heterogeneous Aggregator.}
In this section, we present the three different aggregators of the BasConv, i.e.,~basket aggregator, user aggregator, and item aggregator. As showed in \cref{fig:ubi-aggregator}, these aggregators are built upon the following propagation layers.
\begin{itemize}
    \item \textbf{Self-Propagation Layers.} Each type of node has a self-propagation layer within its corresponding aggregator, i.e., basket-self-propagation layer, user-self-propagation layer, and item-self-propagation layer. In order to reduce the complexity of BasConv, we share the self-propagation layer for all types of nodes, denoted as $\delta$. In each aggregator, the $l$-th self-propagation layer $\delta^{(l)}$ retrieves the self information from $l$ layer~($l \geq 0$). For example, the basket-self-propagation can be calculated as
    \begin{equation}
        \delta^{(l)}(\mathbf{e}_u^{(l)}) = \mathbf{e}_u^{(l)}\mathbf{W}_{sp}^{(l)}.
    \end{equation}
    The user and item self-propagation layers can be defined similarly.
    \item \textbf{Interactive Layers.} We further introduce three interactive layers in these aggregators for the user-basket ($\psi$), user-item ($\gamma$), and item-basket ($\eta$) interactions respectively. For example, we have the user-basket interactive layer
    \begin{equation}
    \psi^{(l)}\left(\mathbf{e}_{u}^{(l)}, \mathbf{e}_{b}^{(l)}\right) = \frac{1}{p_u}\mathbf{e}_{u}^{(l)}\odot \mathbf{e}_{b}^{(l)}\mathbf{W}_{ub}^{(l)},
    \end{equation}
    where $\mathbf{e}_u^{(l)}$ and $\mathbf{e}_b^{(l)}$ are interchangeable, $p_u$ is the normalized factor w.r.t. the degree of the corresponding node, and $\mathbf{W}_{ub}^{(l)}$ is a shared trainable $d\times d$ matrix. Similarly, we can define the user-item interactive layer $\gamma^{(l)}$, the item-basket interactive layer $\eta^{(l)}$, and have the corresponding parameter matrix $\mathbf{W}_{ui}^{(l)}$, $\mathbf{W}_{ib}^{(l)}$.
\end{itemize}
On top of these propagation layers, we are able to formally define the three aggregators as follows.


\subsubsection{Basket Aggregator.}
As showed in \cref{fig:basket-agg}, the aggregated information of basket node $b$ at $l$-th layer $\mathbf{h}_{b}^{(l)}$ should be aggregated from both the embeddings of all the items connected with the basket and the corresponding user embedding:
\vspace{-1mm}
\begin{equation}\label{eq:basket aggregator}
    \begin{split}
    \mathbf{h}_{b}^{(l)} =& \overbrace{\delta^{(l)}\left(\mathbf{e}_{b}^{(l)}\right)}^{\text{basket-self-propagation}} + \overbrace{\psi^{(l)}\left(\mathbf{e}_{u_b}^{(l)}, \mathbf{e}_{b}^{(l)}\right)}^{\text{user-basket interaction}} \\
    &+\underbrace{\sum_{i\in\mathcal{N}_{i}(b)}\eta^{(l)}\left(\mathbf{e}_{b}^{(l)}, \mathbf{e}_{i}^{(l)}\right)}_{\text{item-basket interaction}}.
    \end{split}
\end{equation}
In Eq.~(\ref{eq:basket aggregator}), the aggregated information at the $l$-th layer for basket $b$ is denoted as $\mathbf{h}_{b}^{(l)}$, which is aggregated from the basket-self-propagation layer $\delta$, the user-basket interactive layer $\psi$ and the basket-item interactive layer $\eta$. Each basket connects with only one corresponding user, thus only one basket-associated user $u_b$'s embedding $\mathbf{e}_{u_b}^{(l)}$ passes into the user-basket interactive layer $\psi$ along with basket embedding $\mathbf{e}_{b}^{(l)}$. The basket aggregator collects all the items connected with the basket $b$, which is denoted as the item neighbor function $\mathcal{N}_{i}(b)$. The basket-item interactive layer computes the interactive information of each item with the basket. Summing up the interaction for all the items with the current basket aggregates the item semantics for the basket. 
The aggregated information is passed to an activation function to output the embedding of basket $b$ at $(l+1)$-th layer as 
\begin{equation}
    \mathbf{e}_{b}^{(l+1)} = \sigma\left(\mathbf{h}_{b}^{(l)}\right).
\end{equation}

\begin{figure*}
    \centering
    \includegraphics[width=0.95\linewidth]{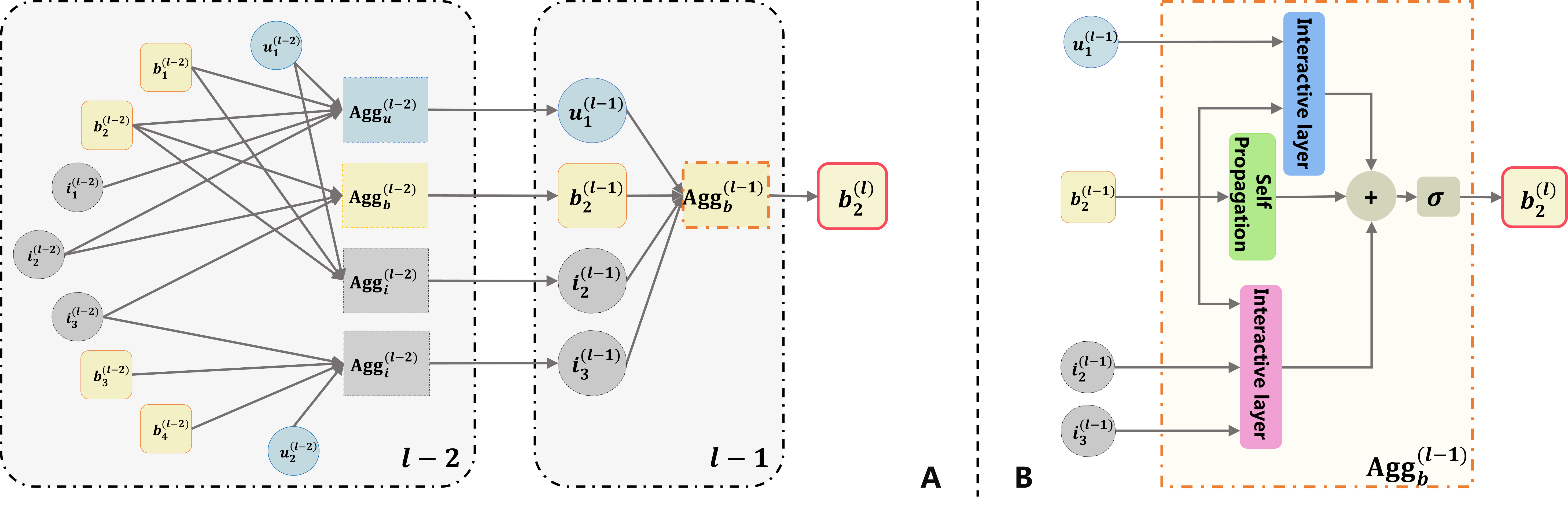}
    \caption{On the left part \textbf{A}, we present the structure of \textbf{BasConv} with an example of aggregating the $l$-th embedding of basket $b_2$, i.e., $\mathbf{b}_2^{(l)}$, based on the UBI example in Fig.~\ref{fig:ubi-aggregator}. In part A, three types of aggregators in the $(l-2)$-th layer aggregate the corresponding embeddings in the $(l-2)$ layer and output the embedding for the $(l-1)$-th layer. Then, the basket aggregator $\mathbf{Agg}_b^{(l-1)}$ produces the embedding $\mathbf{b}_2^{(l)}$ by aggregating the embeddings of $b_2$ and its neighbor nodes in the $(l-1)$-th layer. On the right part \textbf{B}, we show the inner structure of basket aggregator $\mathbf{Agg}_b^{(l-1)}$. The blue block and pink block represent the user-basket and item-basket interactive layers, respectively.}
    \label{fig:basconv_structure}
    \vspace{-1em}
\end{figure*}

The computational flow and structural details for the basket aggregator are presented in \cref{fig:basconv_structure}. We show how to compute the basket embedding $\mathbf{b}_{2}^{(l)}$ by aggregating the information from the previous two layers. First, the aggregators in $(l-2)$-th layer generate the $(l-1)$-th embedding of $b_2$ and its neighbors by aggregating the embeddings of all corresponding neighbor entities. Then the basket aggregator $\mathbf{Agg}_b^{(l-1)}$ aggregates the embeddings of $b_2$'s neighbors and $b_2$ from previous layer. Finally, $\mathbf{Agg}_b^{(l-1)}$ outputs the $l$-th embedding of $b_2$. We present the structure of the basket aggregator on the right-hand side in Fig.~\ref{fig:basconv_structure}. User embedding $\mathbf{u}_{1}^{(l-1)}$ and basket embedding $\mathbf{b}_{2}^{(l-1)}$ are input together into the user-basket interactive layer. Meanwhile, all the connected items are input along with a basket to the basket-item interactive layer. Then they have summed up with the output from the self-propagation layer and input to activation function $\sigma$ to output the embedding. 

\subsubsection{User Aggregator.}
A user $u$ connects with several baskets, thus the aggregated information of user $u$ at the $l$-th layer, denoted as $\mathbf{h}_{u}^{(l)}$, should be learned by aggregating all the basket's information. Moreover, the user-item interaction is important to capture the user's personalized and item semantic information. Hence, we define user aggregator as:
\begin{equation}\label{eq:user aggregator}
    \begin{split}
    \mathbf{h}_{u}^{(l)}=&\overbrace{\delta^{(l)}\left(\mathbf{e}_{u}^{(l)}\right)}^{\text{user-self-propagation}} + \overbrace{\sum_{b\in\mathcal{N}_{b}(u)}\psi^{(l)}\left(\mathbf{e}_{u}^{(l)}, \mathbf{e}_{b}^{(l)}\right)}^{\text{user-basket interaction}}\\
    &+ \underbrace{\sum_{i\in\mathcal{N}_{i}(u)}\gamma^{(l)}\left(\mathbf{e}_{u}^{(l)}, \mathbf{e}_{i}^{(l)}\right)}_{\text{user-item interaction}}.
    \end{split}
\end{equation}
The user-self-propagation layer $\delta$ propagates the self information from the previous layer to the next layer. Each user $u$ connects with a set of baskets, which is denoted as $\mathcal{N}_{b}(u)$. We sum up all the user-basket interactions from user-basket interactive layer $\psi$ to aggregate the information from baskets to the user. In addition to baskets, user $u$ is also connected with a set of items, denoted as $\mathcal{N}_{i}(u)$. The user-item interactive layer $\gamma$ learns the user-item interactive information, which is aggregated for all the corresponding items to user $u$. Then, the aggregated user information is input into an activation function to generate the user embedding as 
$\mathbf{e}_{u}^{(l+1)} = \sigma\left(\mathbf{h}_{u}^{(l)}\right)$.

\subsubsection{Item Aggregator.}
We follow the same rule as in the basket aggregator and user aggregator in the previous section to learn the item aggregated information:
\begin{equation}\label{eq:item aggregator}
    \begin{split}
    \mathbf{h}_{i}^{(l)}=&\overbrace{\delta^{(l)}\left(\mathbf{e}_{i}^{(l)}\right)}^{\text{item-self-propagation}} + \overbrace{\sum_{u\in\mathcal{N}_{i}(u)}\gamma^{(l)}\left(\mathbf{e}_{u}^{(l)}, \mathbf{e}_{i}^{(l)}\right)}^{\text{user-item interaction}}  \\
    &+ \underbrace{\sum_{b\in\mathcal{N}_{b}(i)}\eta^{(l)}\left(\mathbf{e}_{b}^{(l)}, \mathbf{e}_{i}^{(l)}\right)}_{\text{item-basket interaction}}.
    \end{split}
\end{equation}
Item-self-propagation layer retrieves the information from the embedding of item $i$ in the previous layer. We aggregate all the information from the user-item interactive layer and basket-item interactive layer by incorporating the interaction among the item and the corresponding neighboring users, as well as neighboring baskets, respectively. Embedding is generated by passing the  aggregated information to an activation function as $\mathbf{e}_{i}^{(l+1)} = \sigma\left(\mathbf{h}_{i}^{(l)}\right)$.

\subsubsection{Matrix Form.}
Given parameter matrices $\mathbf{W}_{sp}^{(l)}$, $\mathbf{W}_{ub}^{(l)}$, $\mathbf{W}_{ui}^{(l)}$, $\mathbf{W}_{ib}^{(l)}$, the above three aggregators can be represented in explicit matrix forms. Specifically, we have the following updating rules for the user embedding matrices
\begin{equation}
\begin{split}
    \mathbf{E}_{u}^{(l+1)}=&\sigma\Big(
        \mathbf{E}_{u}^{(l)}\mathbf{W}_{sp}^{(l)} + \Tilde{\mathbf{R}}_{ub}\mathbf{E}_{b}^{(l)}\odot \mathbf{E}_{u}^{(l)}\mathbf{W}_{ub}^{(l)} \\
    &+ \Tilde{\mathbf{R}}_{ui}\mathbf{E}_{i}^{(l)}\odot \mathbf{E}_{u}^{(l)}\mathbf{W}_{ui}^{(l)}
    \Big),
\end{split}
\end{equation}
the item embedding matrices
\begin{equation}
\begin{split}
    \mathbf{E}_{i}^{(l+1)}=&\sigma\big(\mathbf{E}_{i}^{(l)}\mathbf{W}_{sp}^{(l)} + \Tilde{\mathbf{R}}_{bi}^{\top}\mathbf{E}_{b}^{(l)}\odot \mathbf{E}_{i}^{(l)}\mathbf{W}_{bi}^{(l)} \\
    &+ \Tilde{\mathbf{R}}_{ui}^{\top}\mathbf{E}_{u}^{(l)}\odot \mathbf{E}_{i}^{(l)}\mathbf{W}_{ui}^{(l)}\big),
\end{split}
\end{equation}
and the basket embedding matrices
\begin{equation}
\begin{split}
    \mathbf{E}_{b}^{(l+1)}=&\sigma\big(\mathbf{E}_{b}^{(l)}\mathbf{W}_{sp}^{(l)} + \Tilde{\mathbf{R}}_{ub}^{\top}\mathbf{E}_{u}^{(l)}\odot \mathbf{E}_{b}^{(l)}\mathbf{W}_{ub}^{(l)}\\
     &+ \Tilde{\mathbf{R}}_{bi}\mathbf{E}_{i}^{(l)}\odot \mathbf{E}_{b}^{(l)}\mathbf{W}_{bi}^{(l)}\big).
\end{split}
\end{equation}
$\Tilde{\mathbf{R}}$ is the normalized matrix of previously defined interaction matrix $\mathbf{R}$ in Sec.~\ref{sec:preliminary}, i.e., $\Tilde{\mathbf{R}}=\mathbf{D}^{-1}\mathbf{R}$ and $\mathbf{D}$ is the corresponding diagonal degree matrix. For example, $\Tilde{\mathbf{R}}_{ub}$ is a user-basket interaction matrix, and $\Tilde{\mathbf{R}}_{ub}=\mathbf{D}_{ub}^{-1}\mathbf{R}_{ub}$  where $D_{ii}=\sum _{j=1}^{|B|}\Tilde{\mathbf{R}}_{ub}(i,j)$.

\begin{table*}[hbt!]
\centering
\vspace{-1em}
\caption{Dataset Statistics}
\vspace{-1em}
\label{tab:data stats}
\begin{tabular}{ccccccc}
\toprule
 \textbf{Dataset} & \textit{\#User} & \textit{\#Item}  & \textit{\#Basket} & \textit{Avg. Basket} & \textit{Avg. Size} & \textit{\#Interactions}\\ 
\midrule
  \textbf{Instacart} & $22,168$ & $40,044$ & $65,672$ & $2.96$ & $37.0$ & $2,495,695$ \\
  \textbf{Walmart} & $44,218$ & $77,599$ & $130,707$ & $2.96$ &$52.5$ & $6,997,572$ \\
\bottomrule
\end{tabular}
\end{table*}

\subsection{Model Prediction.}
BasConv outputs the embeddings of users, baskets, and items after $L$-layer graph convolutions. 
We concatenate embeddings from all layers to incorporate the information from neighborhoods as well as high-order interactions.
For example, the output embedding of basket $b$ is $\mathbf{e}_{b}^{*} = \mathbf{e}_{b}^{(0)} \| \mathbf{e}_{b}^{(1)} \| \cdots \| \mathbf{e}_{b}^{(L)} $, where $\|$ stands for the concatenation operation. 
The same concatenation is applied for both
user embeddings $\mathbf{e}_{u}^{*}$ and item embeddings $\mathbf{e}_{i}^{*}$. With the embedding, we can provide recommendations for the basket $b$ with the candidate items based on their predicted preference scores. For an item $i$, this prediction is defined as
\begin{equation}
    \hat{{y}}(b,i) = \mathbf{e}_{u_b}^{*\top}\mathbf{e}_{i}^{*} + \mathbf{e}_{b}^{*\top}\mathbf{e}_{i}^{*}.
\end{equation}
The first term captures the user-item signals, recommending the items of the corresponding user's interests. The second term models the relationship between the intent of the basket and the embedding of the item. 

\subsection{Optimization.}
We optimize the model based on BPR loss~\cite{rendle2009bpr}. We sample a positive item $i$ and a negative item $j$ for a partially given basket $b$, where the positive item is sampled within the basket and the negative item is sampled from items outside the basket. The final BPR loss is:
\begin{equation}
    \mathcal{L} = -\sum_{(b,i,j)\in \mathcal{S}}\log\sigma\big(\hat{{y}}(b,i) - \hat{{y}}(b,j)\big) + \lambda\|\Theta\|_{2}^{2},
\end{equation}
where the first term denotes the BPR interaction loss, and the second term denotes the regularization to the trainable parameters ($\lambda$ is the regularizing factor). The trainable parameters consist of the embedding parameters and the aggregator parameters. We use Xavier~\cite{glorot2010understanding} initialization. Note for the embedding parameters, we only train the initial user embedding and the item embedding, i.e., only $\{\mathbf{E}_{u}^{(0)}, \mathbf{E}_{i}^{(0)}\}$ are trainable. The initial basket embedding matrix is fixed with all zeros (i.e.,~$\mathbf{E}_{b}^{(0)}=\mathbf{0}$) because of its extremely large number. 
Our models are trained in Tensorflow with the batch-wise Adam~\cite{kingma2015adam} optimizer.

\section{Experiment}

\subsection{Datasets.}
We conduct experiments on two real-world datasets, the \textit{Instacart} dataset\footnote{\url{https://www.instacart.com/datasets/grocery-shopping-2017}} and the dataset collected from the \textit{Walmart} online grocery shopping website\footnote{\url{https://grocery.walmart.com/}}.

\begin{itemize}
    \item \textbf{Instacart} is an online grocery shopping dataset, which is published by \textit{instacart.com}~\cite{instacart}. It contains over 3 million grocery transaction records from over 200 thousand users on around 50 thousand items.  
    \item \textbf{Walmart Grocery} is an online service provided by \textit{walmart.com} for shopping groceries. We sampled  $100$ thousand users, whose transaction data are retrieved to conduct the experiment. 
\end{itemize}
We filter transactions based on their basket sizes to fulfill the requirement of adequate basket signals. Baskets with less than $30$ items and $40$ items for \textit{Instacart} dataset and \textit{Walmart} dataset (respectively) are removed. The statistics of the preprocessed datasets are summarized in Table~\ref{tab:data stats}.

\subsection{Experimental Settings}
\subsubsection{Evaluation Metrics.} We recommend items to complete the partially given baskets. We evaluate the performance of models by the Top-$K$ recommendation metrics, i.e., the  \textit{Recall@K}, \textit{HR@K}, and \textit{NDCG@K}~\cite{he2017neural,wang19neural}. The default $K$ is $100$.
\subsubsection{Baselines.}
We compare our model with the following methods in recommender systems:
\begin{itemize}
    \item \textbf{ItemPop}: From some previous work~\cite{wan2018representing,bai2019personalized}, in real-world, users always buy some popular items. Thus, we design a baseline \textit{itemPop} that considers the user-wise \textit{frequency} of items in the training data as the ranking criteria for recommendation.
    \item
    \textbf{BPR-MF}~\cite{rendle2009bpr}: This is a standard method of modeling user-item interactions with bayesian personalized ranking (BPR) loss. We merge the baskets w.r.t. the same user and sample positive and negative items. We recommend items within baskets based on the corresponding user embeddings and item embeddings. 
    \item
    \textbf{GC-MC}~\cite{berg2017graph}: It is the recent GCNN model to complete the rating matrix. We adopt the idea in~\cite{berg2017graph} using the GCN model~\cite{kipf17semi} to complete the UBI graph. We recommend baskets with items based on the corresponding user embedding. We use one GCN layer connected by an MLP layer structure for predicting the links. 
    \item
    \textbf{NGCF}~\cite{wang19neural}: This is a state-of-the-art GCNN method that explicitly models the high-order interaction of users and items with a multi-layer neural network. However, it has no distinction for the heterogeneous interactions.
    
    \item
    \textbf{Triple2vec}~\cite{wan2018representing}: It is one of the most recent works to address the within-basket recommendation task. We sample the triplets from baskets and train the user and item embeddings. 
    
\end{itemize}

\subsubsection{Parameter Settings.} For a fair comparison, the embedding size is set to $64$ across different methods. The hyper-parameters of BasConv are selected based on the recommendation performance (\textit{Recall}) on the validation set. The learning rate is selected from $\{10^{-5}, 5\times10^{-5}, 10^{-4}, 5\times10^{-4}, 10^{-3}, 5\times10^{-3}\}$. Our model converges best when learning rate is $5\times10^{-4}$ and $10^{-3}$ for the Instacart and Walmart dataset, respectively. The layer size is selected from \{1,2,3,4\} for NGCF and BasConv.

\subsection{Within-Basket Recommendation}
\subsubsection{Overall Comparison.}
In this section, we compare the performance of different models on the within-basket recommendation task. We split $80\%$ items of each basket as training data and the remaining $20\%$ as test data for both of the datasets. All the models are trained on the training data and the hyperparameters are tuned based on the validation data, which is $20\%$ randomly masked data from the training data. We report the within-basket recommendation results on the test data in Table~\ref{tab:within-basket}. The highest value is in bold, and the second-highest value is with a star($*$).

\begin{table} 
\centering
\vspace{-1em}
\caption{Within-Basket Recommendation Comparison}
\vspace{-1em}
\label{tab:within-basket}
\begin{tabular}{clccc}
\toprule
 \textbf{Data}& \textbf{Method} & \textbf{Recall} & \textbf{NDCG}  & \textbf{HR} \\ 
\midrule
  \multirow{7}{*}{Insta.} & ItemPop & $0.1490$ & $0.1604$ & $0.5704$  \\
    & BPR-MF & $0.1687$ & $0.1800$ & $0.6380$ \\
    & triple2vec & $0.1711$ & $0.1855$ & $0.6543$ \\
    & GC-MC & $0.1758$ & $0.1871$ & $0.6529$ \\
    & NGCF & $0.1887*$ & $0.2013*$ & $0.71729*$ \\
    & BasConv & $\textbf{0.2092}$ & $\textbf{0.2281}$ & $\textbf{0.7712}$ \\
    \midrule
    & Improv.\% & $\textbf{7.34\%}$ & $\textbf{1.74\%}$ & $\textbf{9.89\%}$ \\
    \midrule
    \midrule
  \multirow{6}{*}{Wal.} & ItemPop & $0.0490$ & $0.0586$ &  $0.2732$  \\
    & BPR-MF & $0.0430$ & $0.0620$ & $0.2619$ \\
    & triple2vec & $0.0462$ & $0.0663$ & $0.2713$ \\
    & GC-MC & $0.0392$ & $0.0706$  & $0.2874$ \\
    & NGCF & $0.0492*$ & $0.0722*$ & $0.2903*$  \\
    & BasConv & $\textbf{0.0530}$ & $\textbf{0.0841}$ & $\textbf{0.3394}$ \\
    \midrule
    & Improv.\% & $\textbf{7.72\%}$ & $\textbf{16.48\%}$ & $\textbf{16.91\%}$ \\
    
    
\bottomrule
\end{tabular}
\end{table}

BasConv improves the Recall, NDCG, and Hit Ratio $7.72\%$~($7.34\%$), $15.9\%$~($1.74\%$) and $18.60\%$~($9.89\%$) respectively on the Walmart~(Instacart) dataset. The improvement comes from two aspects: (1) BasConv aggregates the information from high-order structure via heterogeneous aggregators, which capture multi-type node information. (2) The heterogeneous interactive layers explicitly model the multi-type interactions in the UBI graph. The results prove that our proposed BasConv model can retrieve the high-order collectivity and the heterogeneity pattern from the UBI graph, and hence improve the performance of recommendation.

Although user-wise item popularity (ItemPop) lacks generalization power, as it is able to memorize users' simple shopping patterns, it still 
on both datasets as it explicitly models the interactions of user-items. However, since all existing methods ignore the collectivity and heterogeneity of the basket recommendation problem, they can be outperformed by 
BasConv in terms of all three evaluation metrics.

\subsubsection{Sensitivity Analysis.} 
\begin{figure*}[hbt!]
\begin{subfigure}{.33\textwidth}
    \centering
    \includegraphics[width=\textwidth]{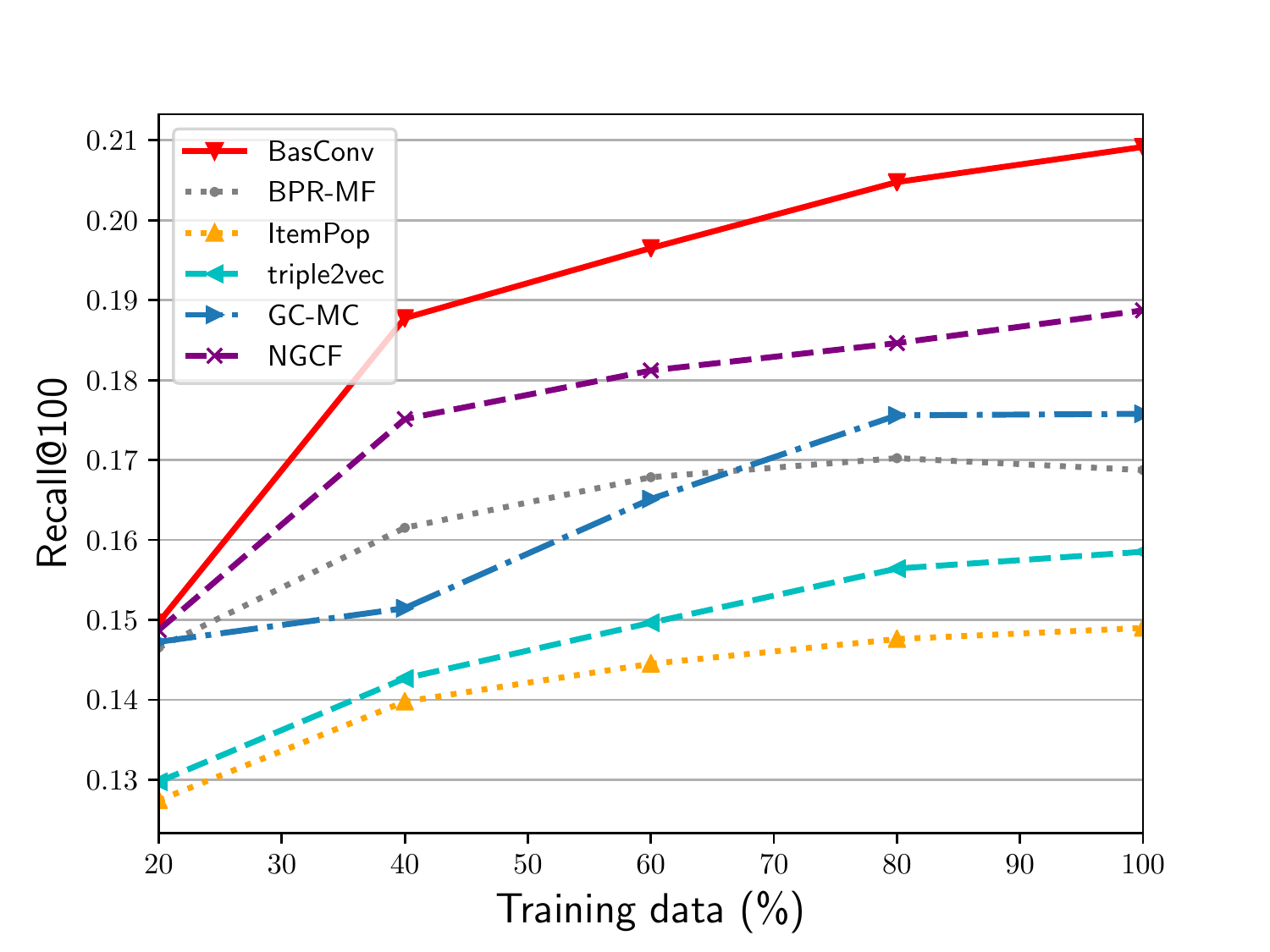}
    \caption{\text{Recall on Instacart}}
    \label{fig:recall_inscart}
\end{subfigure}\hfill
\begin{subfigure}{.33\textwidth}
    \centering
    \includegraphics[width=\textwidth]{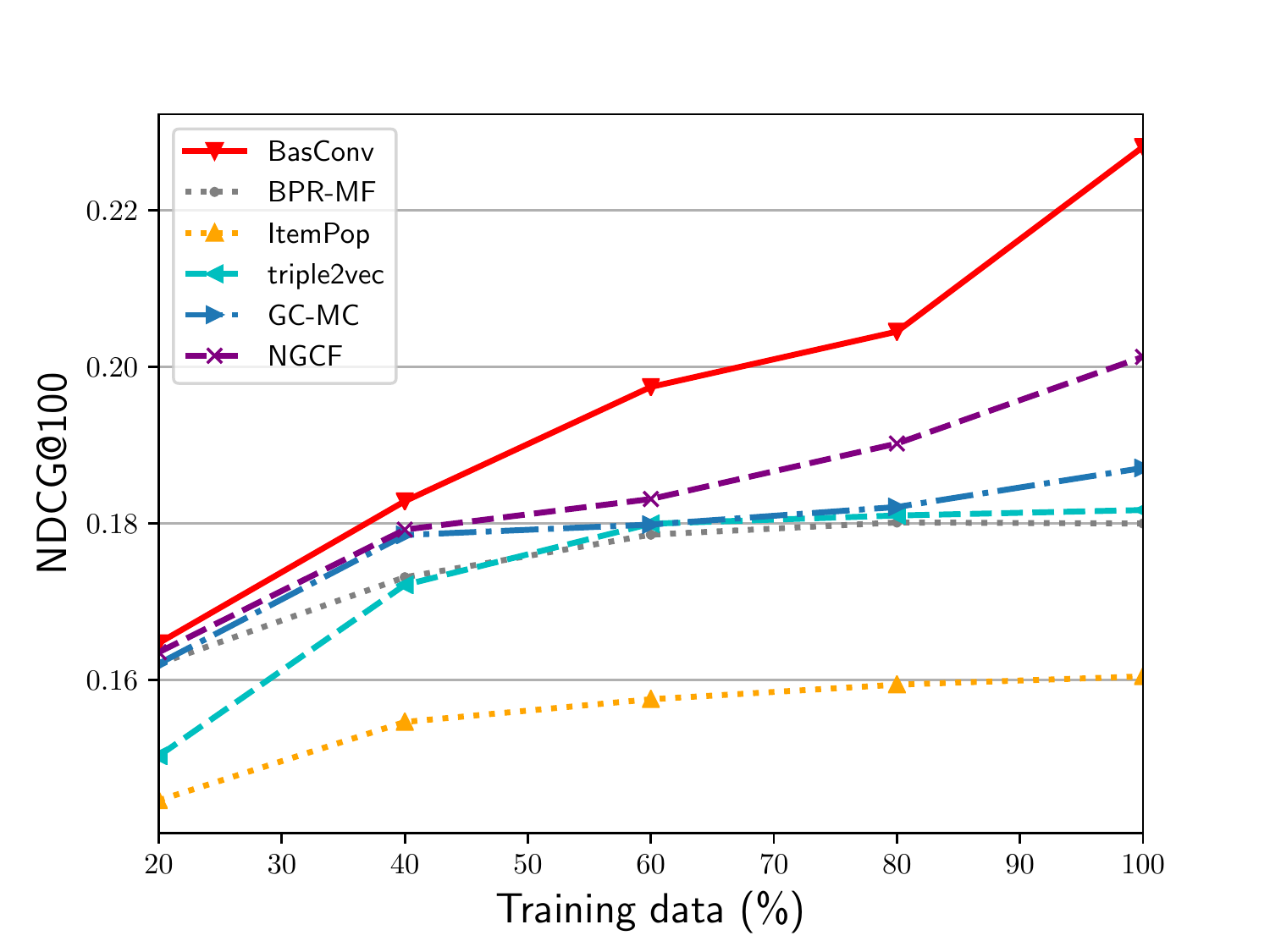}
    \caption{\text{NDCG on Instacart}}
    \label{fig:NDCG_inscart}
\end{subfigure}\hfill
\begin{subfigure}{.33\textwidth}
    \centering
    \includegraphics[width=\textwidth]{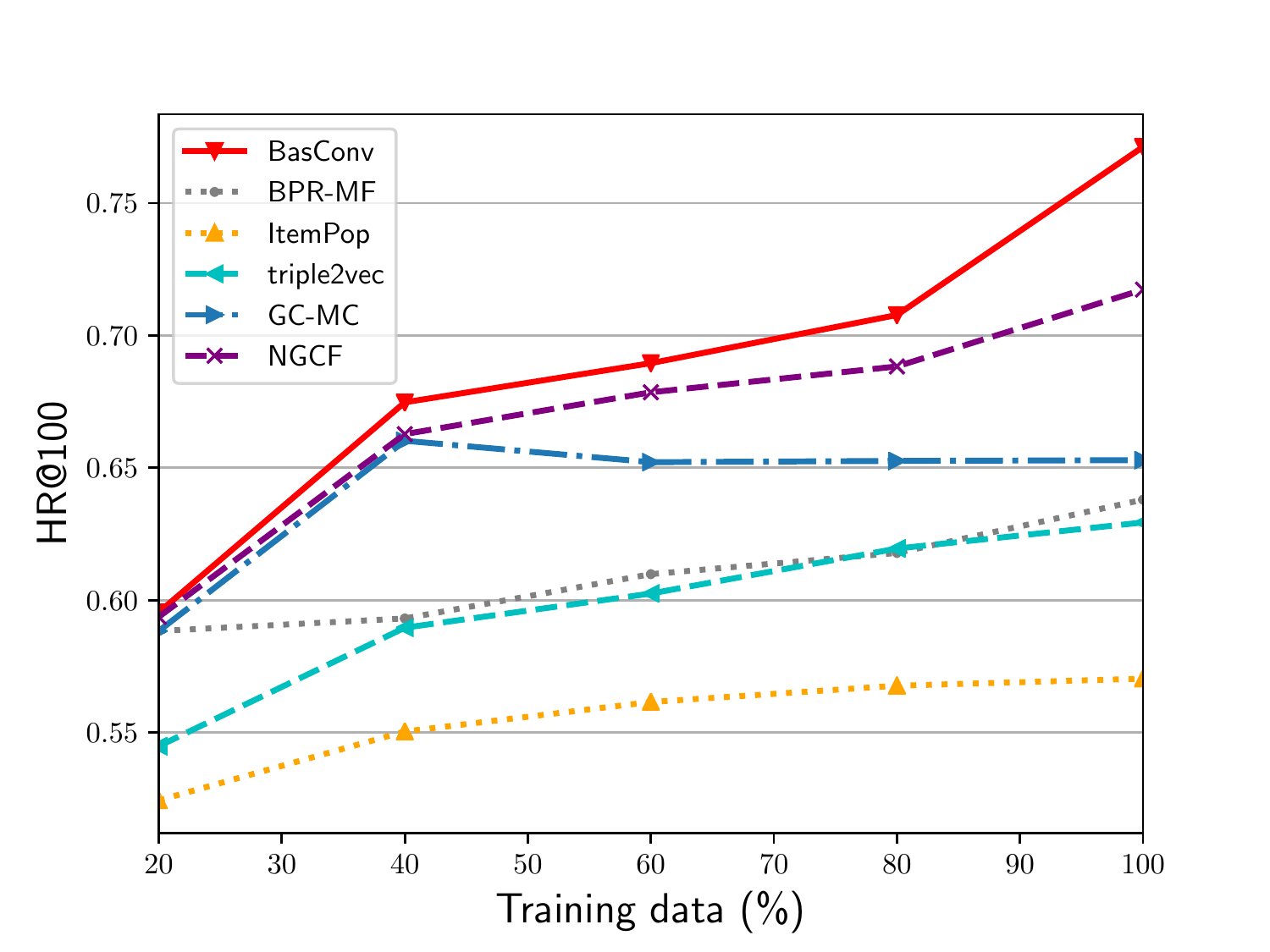}
    \caption{\text{Hit Ratio on Instacart}}
    \label{fig:HR_inscart}
\end{subfigure}
\hfill
\\
\begin{subfigure}{.33\textwidth}
    \centering
    \includegraphics[width=\textwidth]{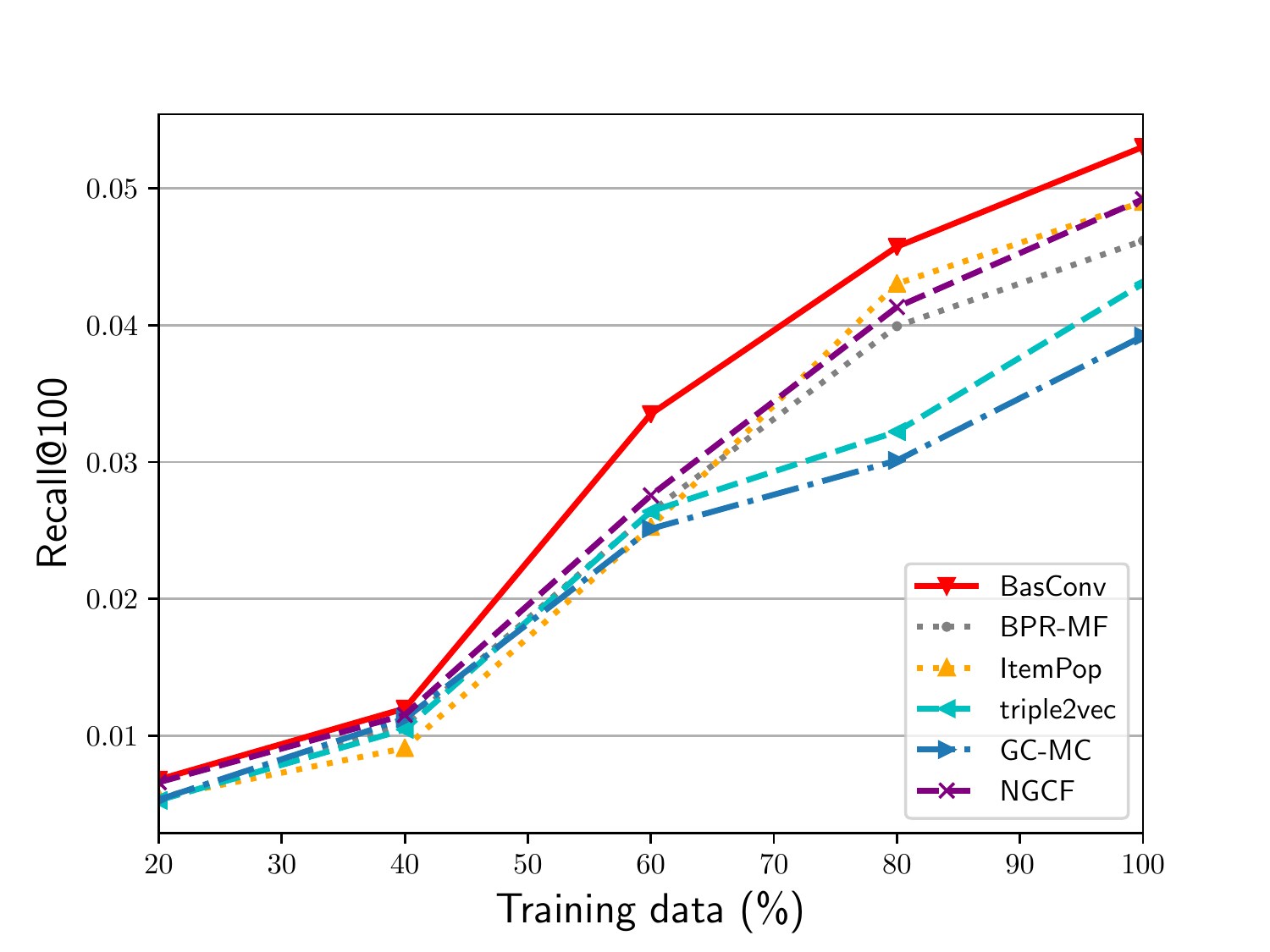}
    \caption{\text{Recall on Walmart}}
    \label{fig:recall_walmart}
\end{subfigure}
\begin{subfigure}{.33\textwidth}
    \centering
    \includegraphics[width=\textwidth]{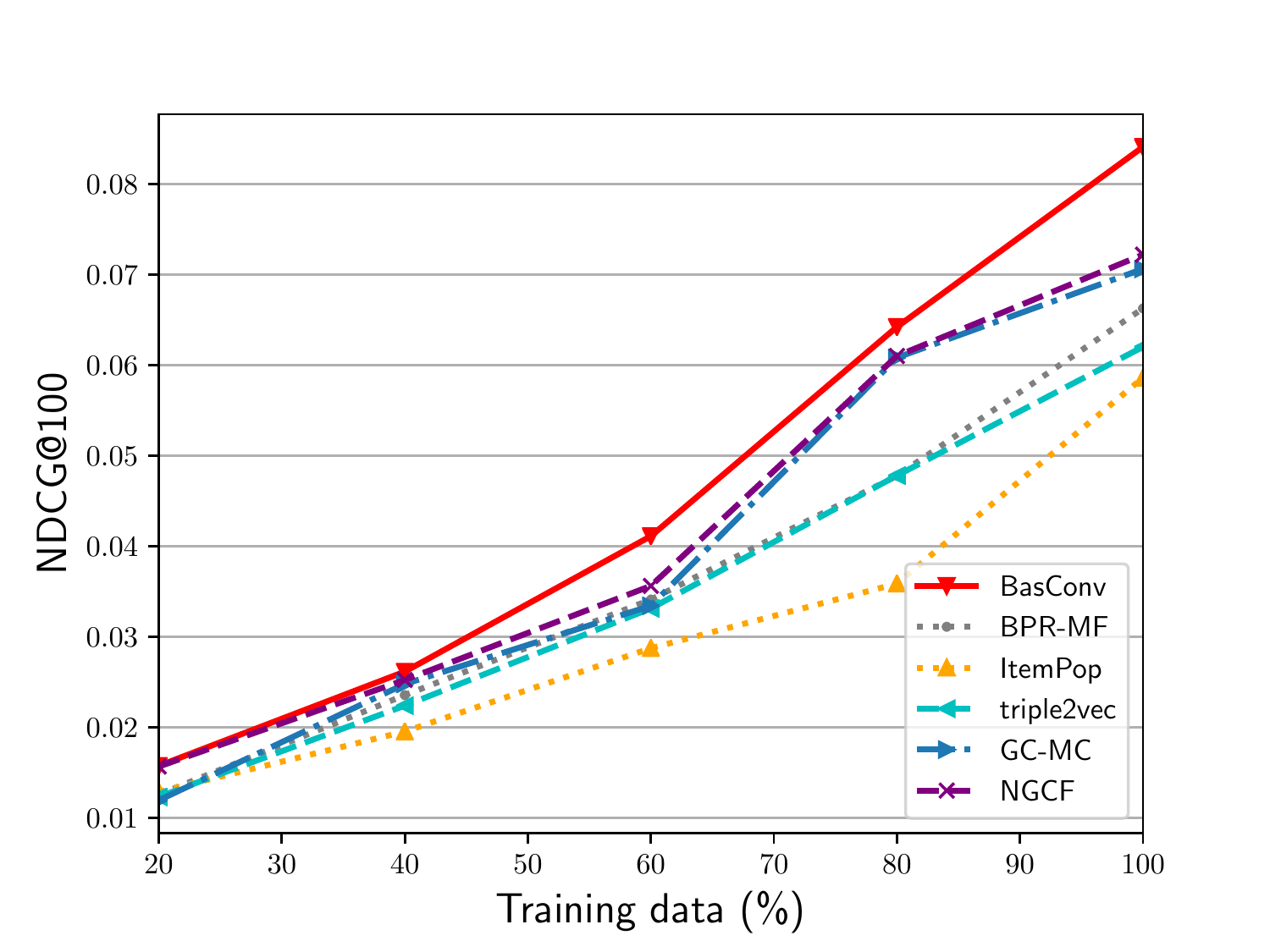}
    \caption{\text{NDCG on Walmart}}
    \label{fig:NDCG_walmart}
\end{subfigure}
\begin{subfigure}{.33\textwidth}
    \centering
    \includegraphics[width=\textwidth]{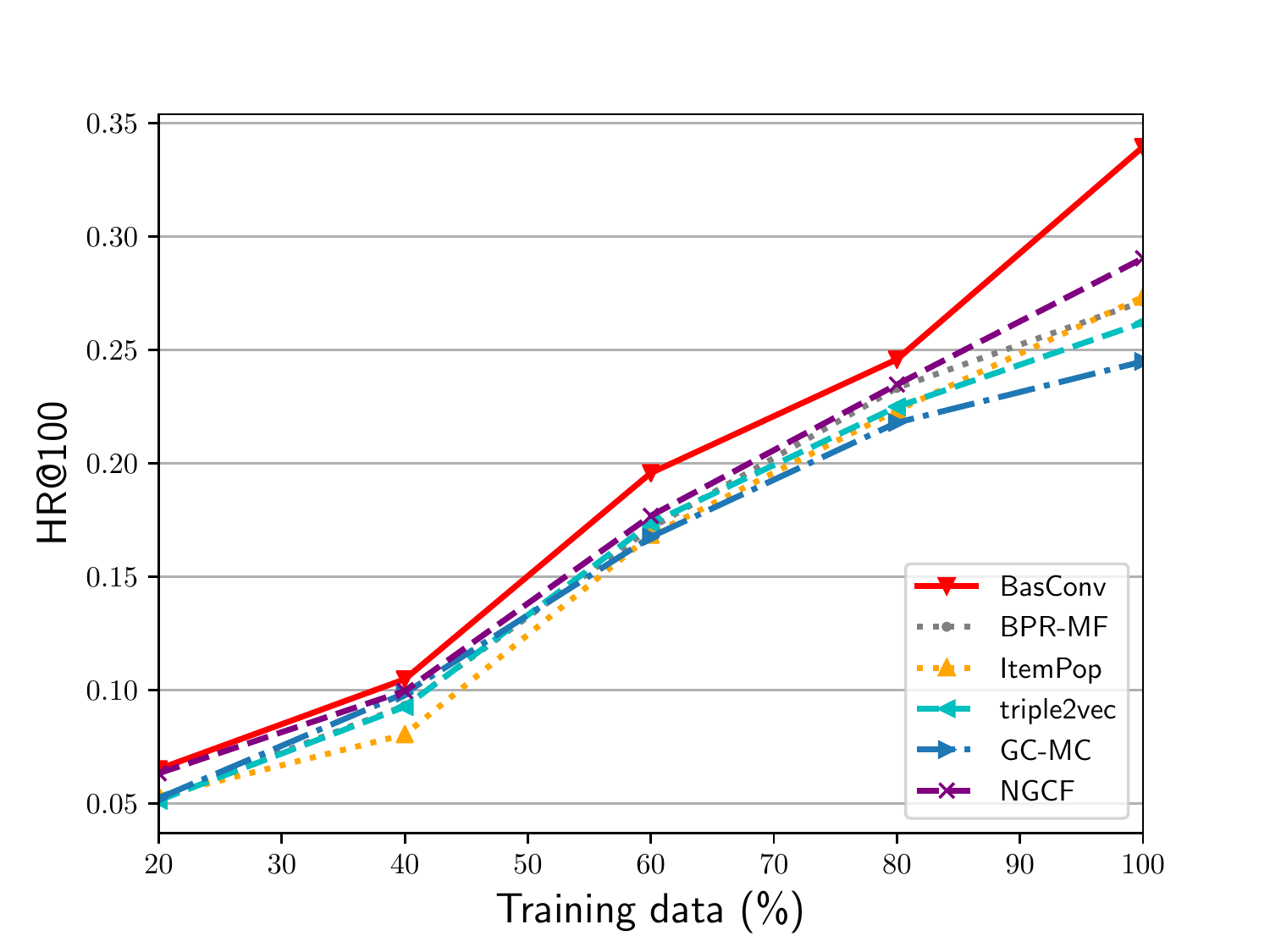}
    \caption{\text{Hit Ratio on Walmart}}
    \label{fig:HR_walmart}
\end{subfigure}
\caption{Model Performance w.r.t. different percentage of training data on two datasets.}
 \vspace{-1em}
\label{fig:sensitivity_analysis}
\end{figure*}

Data sparsity can spoil the performance of recommender systems since using limited
interactions might be 
insufficient to comprehensively retrieve entity semantics. 
We thus investigate the sensitivity of models w.r.t. the number of interactions of basket-items. We compare BasConv with other methods when training data is sampled $\{20\%, 40\%, 60\%, 80\%, 100\%\}$ from the training data used in the previous section to study the sensitivity of different methods. The results are presented in Fig.~\ref{fig:sensitivity_analysis}. We observe that BasConv consistently outperforms other methods as the data volume increases, which proves that high-order collectivity and heterogeneity is important to retrieve the semantics of the UBI graph. Also, we find that all methods have a tendency to perform better when data size increases, but BasConv improves faster than other models. 

\begin{table}
\centering
\vspace{-1em}
\caption{Effects of Layer Number}
\vspace{-1em}
\label{tab:layer_size}
\begin{tabular}{clccc}
\toprule
 \textbf{Data}& \textbf{Method} & \textbf{Recall} & \textbf{NDCG}  & \textbf{HR} \\ 

\midrule
  \multirow{4}{*}{Insta.} & BasConv-1 & $0.1799$ & $0.2221$ & $0.7666$  \\
    & BasConv-2 & $0.1818$ & $0.2235$ & $0.7699$ \\
    & BasConv-3 & $\textbf{0.2092}$ & $\textbf{0.2281}$ & $\textbf{0.7712}$ \\
    & BasConv-4 & $0.1868$ & $0.2238$ & $0.7605$ \\
    \midrule
    \midrule

   \multirow{4}{*}{Wal.} & BasConv-1 & $0.0499$ & $0.0728$ & $0.2892$  \\
    & BasConv-2 & $\textbf{0.0530}$ & $\textbf{0.0841}$ & $\textbf{0.3394}$ \\
    & BasConv-3 & $0.0500$ & $0.0837$ & $0.3305$ \\
    & BasConv-4 & $0.0496$ & $0.0830$ & $0.3272$
    \\
    
    
\bottomrule
\end{tabular}
\end{table}

\subsection{Study of BasConv.}
BasConv has a multi-layer structure which collects the high-order interactions in the UBI graph. The layer number is an important hyper-parameter for BasConv. We conduct experiments on a different number of layers of BasConv to investigate the model performance change. The layer number is chosen from \{1,2,3,4\}, and the results are displayed in Table~\ref{tab:layer_size}. We find that on the Instacart dataset, BasConv performs best when layer number is $3$, while on Walmart dataset, BasConv performs best when the layer size is $2$. A possible reason for this performance drop could be higher-order interactions may contain less relevant information thus introducing noises in the model.

\subsection{Case Study.} In this section, we show the ranking results of BasConv in Table~\ref{tab:case_study} for a real-world case study\footnote{We delete some irrelevant descriptive words to save space.}. 
In this table, we show a portion of the training items in the given basket at the top, the test items (i.e.,~the ground truth) in the same basket in the middle, and the recommended items at the bottom. 
We observe the following two potential intents in the given basket: (1) the `Lemon Verbena Dish Soap' and the `Sweeper Dry Sweeping Cloth Refills' may indicate the user's intention of \textit{cleaning} and (2) the  `Homestyle Belgian Waffles' may indicate the user's intention of \textit{breakfast}.
BasConv can identify both of them and recommends \textit{milk} and \textit{eggs} for the breakfast intention and \textit{dishwasher detergent} for cleaning, which are all verified by 
the ground truth. 
We have successfully predicted 3 products in the top 7 recommendations while others preserve reasonable explanations.
For instance, the `Brioche Slider Buns' is also related to the breakfast purpose and `Pasta' is recommended as `Cheddar' is in the given basket. This real-world use case justifies that BasConv is able to identify the intent of the basket, model the semantics of items, and thus provide a satisfying recommendation.




\begin{table}[hbt!]
\centering
\vspace{-1em}
\caption{A case study on instacart dataset. We find two different intents in the given basket---\textit{cleaning} (in {\color{red} red}) and \textit{breakfast} (in {\color{blue} blue}). The correctly predicted items are \underline{underlined}.}
\vspace{-0.5em}
\label{tab:case_study}
\resizebox{\linewidth}{!}{
\begin{tabular}{c}
\toprule

\makecell[c]{\textbf{Items within the partially given basket}\\ \textbf{(Condition)}}\\
\midrule
{\color{red} Lemon Verbena Dish Soap}\\
{\color{red} Sweeper Dry Sweeping Cloth Refills} \\
{\color{blue} Homestyle Belgian Waffles} \\
Lightly Salted Kettle Potato Chips - Sea Salt \\
{Baked Rice and Corn Puffs, Aged White Cheddar} \\
Chocolate Brownie Kid Z Bar \\
...\\
\midrule\midrule
\makecell[c]{\textbf{Test items purchased in the same basket}\\ \textbf{(Ground Truth)}}\\
\midrule
{\color{blue} Grade A Large Brown Eggs} \\
{\color{blue} Unsweetened Original Milk} \\
Gala Apples \\
{\color{red} Scent Dishwasher Detergent} \\
{\color{red} Free \& Gentle Fabric Softener Dryer Sheets} \\
{\color{red} Unscented Liquid Laundry Detergent} \\
\midrule\midrule
\makecell[c]{\textbf{Recommended items to the partically given basket}\\ \textbf{(Prediction)}}\\
\midrule
Organic Chocolate Chip ZBar Kids Energy Snack \\
{\color{red} \underline{Scent Dishwasher Detergent}}\\
{\color{blue} Brioche Slider Buns} \\
{Penne Rigate 41 Pasta} \\
{\color{blue} \underline{Unsweetened Original Milk}} \\
{\color{blue} Cage Free 100\% Liquid Egg Whites} \\
{\color{blue} \underline{Grade A Large Brown Eggs}} \\
\bottomrule
\end{tabular}}
\end{table}

\section{Conclusion}
In this paper, we introduced the user-basket-item (UBI) graph where the basket entity represents the intent of the shopping basket. Upon this, we formulated the within-basket recommendation problem as a link prediction problem. In order to solve the high-order collectivity and heterogeneity challenges in identifying basket intent, we introduced three types of aggregators to incorporate heterogeneous interactive signals and collectivity semantics. Extensive experiments on Instacart and Walmart datasets demonstrated the effectiveness of BasConv in modeling high-order collectivity and heterogeneity. The real-world case study validated that our proposed model could identify the intents of the current shopping basket. 
\section{Acknowledgements}
This work is supported in part by NSF under grants III-1526499, III-1763325, III-1909323, and CNS-1930941.

\bibliographystyle{IEEEtran}
\bibliography{reference}
\end{document}